\documentclass[final]{siamltex}

\usepackage{graphicx}
\usepackage{amssymb}
\usepackage{epstopdf}
\usepackage{amsmath}
\title{Manifold learning techniques and model reduction applied to dissipative PDEs}

\author{Benjamin E. Sonday\footnotemark[2]\ \footnotemark[7]
\and Amit Singer\footnotemark[2]\ \footnotemark[3] \footnotemark[8]
\and C. William Gear\footnotemark[4] \footnotemark[5] \footnotemark[8]
\and Ioannis G. Kevrekidis\footnotemark[2] \footnotemark[4] \footnotemark[8]}

\begin{document}
\maketitle
\renewcommand{\thefootnote}{\fnsymbol{footnote}}
\footnotetext[2]{Program in Applied and Computational Mathematics, Princeton University, Princeton, NJ 08544}
\footnotetext[3]{Department of Mathematics, Princeton University, Princeton, NJ 08544}
\footnotetext[4]{Department of Chemical Engineering, Princeton University, Princeton, NJ 08544}
\footnotetext[5]{NEC Laboratories USA, retired}
\footnotetext[6]{Corresponding author ({\tt yannis@princeton.edu})}
\footnotetext[7]{Partially supported by the DOE CSGF (grant number DE-FG02-97ER25308) and NSF GRFP}
\footnotetext[8]{Partially supported by the DOE (grant number DE-SC0002097)}
\renewcommand{\thefootnote}{\arabic{footnote}}

%sdfjklsdfjkl Ben fix headers too
%
%\author{Benjamin E. Sonday\thanks{Program in Applied and Computational Mathematics, Princeton University, Princeton, NJ 08544} \and Amit %Singer\thanks{Program in Applied and Computational Mathematics and Department of Mathematics, Princeton, NJ 08544} \and C. William Gear\thanks{Department %of Chemical Engineering, Princeton University, Princeton, NJ 08544 and NEC Laboratories USA, retired} \and Ioannis G. Kevrekidis\thanks{Program in %Applied and Computational Mathematics and Department of Chemical Engineering, Princeton, NJ 08544 ({\tt yannis@princeton.edu})}}

%\author{Paul Duggan\thanks{Composition Department, Society
%        for Industrial and Applied Mathematics, 3600 Univeristy 
%        City Science Center, Philadelphia, Pennsylvania, 
%        19104-2688 ({\tt duggan@siam.org}).}
%        \and Various A.~U. Thors\thanks{Various Affiliations, 
%        supported by various foundation grants.}}

%\begin{document}

%\maketitle

%\author[label2,label6]{Benjamin Sonday}
%\author[label2,label5]{Ioannis G. Kevrekidis}

%\fntext[label6]{Corresponding author:  bsonday@math.princeton.edu}
%\address[label2]{Program in Applied and Computational Mathematics, Princeton University, Princeton, NJ 08544}
%\address[label5]{Department of Chemical Engineering, Princeton University, Princeton, NJ 08544, USA}

\begin{abstract}
We link nonlinear manifold learning techniques for data analysis/compression
with model reduction techniques for evolution equations with time scale separation.
In particular, we demonstrate a ``nonlinear extension" of the POD-Galerkin
approach to obtaining reduced dynamic models of dissipative evolution equations.
The approach is illustrated through a reaction-diffusion PDE, and the performance
of different simulators on the full and the reduced models is compared.
We also discuss the relation of this nonlinear extension with the so-called
{\it nonlinear Galerkin} methods developed in the context of
Approximate Inertial Manifolds.
%
%Our discussion and comparisons are based on a reaction-diffusion illustrative example.
\end{abstract}


\pagestyle{myheadings}
\thispagestyle{plain}
\markboth{B.~E. SONDAY, A. SINGER, C.~W. GEAR, AND I.~G. KEVREKIDIS}{B.~E. SONDAY, A. SINGER, C.~W. GEAR AND I.~G. KEVREKIDIS}

\section{Introduction}

The purpose of this paper is to extend established model reduction
methods for large-scale dynamical systems characterized by separation
of time scales by linking them to recently developed (nonlinear)
manifold learning techniques--in particular, with the diffusion map (DMAP) approach of Coifman and coworkers
\cite{coifman1,coifman2}; see also \cite{belkin2003}.

The general setting involves a high-dimensional, \textit{stiff} system of differential equations of the form
\begin{equation} \label{mainEq}
 \frac{d \vec y}{dt} = \vec f(\vec y)
\end{equation}
with $\vec f:\,\,\mathbf{R}^d \longrightarrow \mathbf{R}^d$.
We will focus on problems where this system arises in the context of
discretizing a dissipative partial differential equation:
an equation of the form
\begin{equation}\label{eq:AIM}
\frac{\partial}{\partial t}u + Au = F(u),
\end{equation}
where $A$ is a positive self-adjoint operator with a discrete spectrum
and $F(u)$ is a ``well-behaved'' function of $u$ \cite{initmanbook,constantin1985attractors,foias1988inertial,NLGreview,garcia1998postprocessing,titi1990approximate}.
We chose this type of example because we will later draw some analogies between our
approach and the so-called {\it nonlinear Galerkin} reduction techniques \cite{foias1989exponential,aimshort,aim,garcia1998postprocessing,titi1990approximate} developed in precisely this
context of Approximate Inertial Manifolds.
We note, however, that the reduction procedure we will discuss can
also be attempted for large systems of ODEs characterized by
separation of time scales (and the associated stiffness) that arise
in other situations--for example, in large, complicated chemical
kinetic networks.

For our prototypical system (equation (\ref{mainEq})), there exists a slow,
attracting, invariant manifold to which trajectories quickly decay.
In the framework of Approximate Inertial Manifolds, this means that
when solutions to equation (\ref{eq:AIM}) are projected onto the
complete set of eigenfunctions of $A$, the long-term behavior of the
solution components in the higher eigenfunctions can be
parameterized by (depicted as a graph of a function over) the
solution components in the leading eigenfunctions.
The result is a low-dimensional manifold in an infinite-dimensional space
(or, for truncated approximations, in a high-dimensional space).
Given a collection of points sampled from such a manifold, we wish to
develop a reduced set of dynamic equations that describe the dynamics
{\em on this manifold}.
Ideally, the dimensionality of this reduced set would be the ``intrinsic''
dimension of the long-term dynamics: the dimension of
the manifold.
Nonlinear Galerkin techniques share this goal, but they are not
based on data sampling on the manifold; they are based instead on
approximating it as a graph of a function above the first few eigenfunctions of the operator (or eigenvectors
in the case of its discretization).
What we propose here is an extension of what is
referred to as ``the POD-Galerkin approach"--sampling data on the manifold, using
Principal Component Analysis (PCA) to compress the data (see, e.g. \cite{jol_pca}), and performing
(various forms of) Galerkin projection of the dynamics on the
leading POD modes to obtain reduced models of the long-term dynamics themselves (see, e.g., \cite{holmes1998turbulence,deane,podg1,podg2,sirovichPOD}).
Since principal component analysis passes optimal (in a well-defined sense) approximating hyperplanes through the
data points, one can easily envision intrinsically low-dimensional (but curved) manifolds that require much higher dimensional hyperplanes to successfully embed them.
Remedying this potential discrepancy between intrinsic manifold dimension and the lowest-dimensional POD-Galerkin that can successfully reproduce the long-term dynamics is the focus of this work.

Modern manifold learning techniques can be thought of as (nonlinear)
extensions of PCA, in the sense that they ``learn" the geometry
of the low-dimensional manifolds on which the data lie, and pass optimal (again in a well-defined sense)
nonlinear approximating manifolds through the data points.
Our procedure utilizes \textit{diffusion maps} (DMAPs) to learn the geometry of the slow manifold from simulation data, and the related \textit{Nystr\"om extension} to rewrite the dynamics of (\ref{mainEq}) exploiting this geometry.
Some interpolation (in the form of the Nystr\"om extension) is required, but it is performed in the reduced-dimension
embedding space  (as opposed to the original, high-dimensional, data space).
Model dimensionality as well as model equation stiffness is thus hopefully reduced, so that both explicit and implicit temporal integration methods may exhibit savings over the original, full model simulation.
The premise is that the ``basis functions'' provided by the DMAP process can be (sometimes significantly)
fewer in number than the basis functions provided by methods such as POD, since they represent the true underlying dimensionality of the slow manifold.

Reducing the model with the help of these ``nonlinear'' modes
can help identify significant components of the dynamics, enhance intuition about the dynamical system behavior,
and faciliate computations.
We will see, however, that the nonlinearity of the reduction technique also gives rise to certain difficulties, which may offset
the benefit of the ``more parsimonious" manifold parametrization.
Other approaches to model reduction can be analytical, such as ``quasi-steady-state''/partial-equilibrium assumptions \cite{pea,qssa}, or computer-assisted, such as the rate-controlled constrained equilibrium method (RCCE) \cite{rcce}, computational singular perturbation (CSP) \cite{csp}, intrinsic low-dimensional manifolds (ILDM) \cite{ildm} (essentially, first-order CSP), functional iteration \cite{fraser}, and even conditions on various derivatives \cite{ptasm,cmalc,curry,kreiss,lorenz,aenio}.

The paper is organized as follows.
We start with a concise description of DMAPs for manifold learning using high-dimensional data.
We then briefly review the POD-Galerkin approach, and present our method as its natural nonlinear analogue.
We then compare our method with nonlinear Galerkin reduction techniques through an illustrative example,
and conclude with a brief summary and discussion of open problems and possible extensions.

\section{Brief introduction to DMAPs}
Diffusion maps have recently emerged as a fast, robust,
nonlinear dimensionality reduction tool
\cite{coifman1,coifman2}; see also \cite{belkin2003}.
Starting with an ensemble of data (e.g. points in a high-dimensional space), diffusion maps (DMAPs)
help determine whether they can be embedded in a lower-dimensional space,
and also find and parametrize a ``best'' (and possibly nonlinear)
low-dimensional manifold that (approximately) contains the data.
In actuality, the only input to the DMAP algorithm is a scalar
\textit{similarity measure} between each pair of entries in the data
ensemble; therefore, DMAPs can also be used on data that are not
necessarily points in Euclidean space \cite{cecilia,frewenchem,physreve}.
In our context, however, the data ensemble does consist of points (represented
as vectors) sampled from ODE trajectories; we therefore present the DMAP
algorithm for data ensembles that consist of vectors $\vec X_i \in \mathbf{R}^d$.
Typically, the similarity measure becomes negligible beyond a {\it local
neighborhood} of each data point; this is related to the notion that
large Euclidean distances may not reliably approximate geodesic
distances on a general nonlinear manifold.
These pairwise similarities are used in the DMAP algorithm to
construct a matrix whose leading eigenvectors \textit{nonlinearly}
embed the data set in a lower-dimensional Euclidean space.
Euclidean distance in the new, reduced space approximates 
the \textit{diffusion distance}, a quantity related to manifold
geodesic distance (for a rigorous definition of the diffusion distance
see \cite{coifman_diffusion}).
The DMAP can thus provide a global manifold parametrization based
on local information only, much like the unfolding of a crumpled
towel to a two-dimensional rectangle.

To construct a low-dimensional embedding for a data set of $M$
individual points (represented as $d$-dimensional real vectors, $\vec X_1$,...,$\vec X_M$)
we start with a similarity measure between each pair of vectors
$\vec X_i$, $\vec X_j$.
The similarity measure is a nonnegative
quantity $W(i,j) = W(j,i)$ satisfying certain additional
``admissibility conditions" \cite{coifman1}.
As a concrete example, consider the Gaussian and Epanechnikov similarity measures
(based on the standard $L^2$ norm):
\begin{equation} \label{eq:Gauss}
W(i,j) = \text{exp} \left[ - \left(\frac{\left\|\vec X_i - \vec
X_j\right\|}{\epsilon} \right)^2 \right]
\end{equation}
and
\begin{equation} \label{eq:Epan}
 W(i,j) = (1-\left\|\vec X_i-\vec X_j\right\|^2/\epsilon^2) \mathbf{1}_{\{\left\|\vec X_i-\vec X_j\right\|<\epsilon \}}.
\end{equation}
A weighted Euclidean norm may be chosen over the standard $L^2$ norm
in situations where the values of different components of $\vec X$
may vary over disparate orders of magnitude.
$\epsilon$ defines a characteristic scale which quantifies the
``locality'' of the neighborhood within which Euclidean distance can
be used as the basis of a meaningful similarity measure
\cite{coifman1}. A systematic approach to determining appropriate
$\epsilon$ values is discussed below.

Next, we define the diagonal matrix $\mathbf{D}$ by
\begin{equation*}
D(i,i) = \sum^M_{k=1} W(i,k),
\end{equation*}
and then we compute the first few right eigenvectors corresponding
to the largest eigenvalues of the stochastic matrix
\begin{equation*}
\mathbf{K} = \mathbf{D}^{-1} \mathbf{W}.
\end{equation*}
In MATLAB, for instance, this can be done with the command
$[\mathbf{V},\mathbf{L}] = \mbox{eigs}(\mathbf{K},n)$, where $n$ is
the number of top eigenvalues we wish to keep (we typically are only
interested in the first few).

This gives a set of real eigenvalues $\lambda_0 \geq \lambda_1 \geq
... \geq \lambda_{n} \geq ... \geq 0$ with corresponding eigenvectors $\{\vec
\psi_j\}_{j=0}^{n}$.
Since $\mathbf{K}$ is stochastic, $\lambda_0 = 1$ and $\vec \psi_0 =
[1\, 1 \, ... \, 1]^T$.
The $n$-dimensional representation of a particular $d$-dimensional
data vector, $\vec X_i$, is given by the \textit{diffusion map}
$\vec \Psi_n:\, \mathbf{R}^d \longrightarrow \mathbf{R}^n$, where
\begin{equation*}
\vec \Psi_n(\vec X_i) = [\lambda_1^t \vec \psi_1^{(i)}, \lambda_2^t \vec \psi_2^{(i)}, ..., \lambda_n^t \vec
\psi_n^{(i)}],
\end{equation*}
a mapping which is only defined on the $M$ recorded data vectors.
Here, $t$ represents the ``diffusion time''; to keep things simple, we choose $t=1$.
In other words, the vector $\vec X_i$ is mapped to a vector whose
first component is the $i$th component of the first nontrivial
eigenvector, whose second component is the $i$th component of the
second nontrivial eigenvector, etc.
If a gap in the eigenvalue spectrum is observed
between eigenvalues $\lambda_n$ and $\lambda_{n+1}$, then $\vec \Psi_n$
may provide a useful low-dimensional representation of the data
set \cite{belkin2003,coiffp}.
In fact, when this gap is present (and when the DMAP is scaled as $\vec \Psi_n(\vec X_i) = [\lambda_1 \vec \psi_1^{(i)},
\lambda_2 \vec \psi_2^{(i)}, ..., \lambda_n \vec \psi_n^{(i)}]$), Euclidean distance in the DMAP space of
only these first $n$ eigenvectors will accurately
approximate the {\em diffusion distance} mentioned above\footnote{It should be noted that sometimes, $\vec \psi_i$ and $\vec \psi_j$ (for some $i<j \leq n$)
contain redundant information.
Consider, for instance, a
two-dimensional rectangular sheet (perhaps in a high-dimensional ambient
space) of long length and narrow width, for which we wish to obtain a two-dimensional
parametrization.
We would therefore desire only two eigenvectors (eigenfunctions)
of diffusion, one corresponding to the coordinate in the width direction and one
to the coordinate in the length direction.
However, many of the computed eigenvalues for diffusion in the ``long" direction may be greater than the
first eigenvalue whose eigenvector corresponds to diffusion in the ``width"  direction.
We may then want to ignore, in our embedding, many of the eigenvectors  corresponding to
the long direction.
Measures such as mutual information have been suggested to detect
such ``redundant" eigenvectors \cite{amit_conv}.}.
%
%It is also interesting to note that if the data come from a Markovian stochastic
%process, the eigenvectors and eigenvalues are approximations to the
%eigenfunctions and eigenvalues of the corresponding backward
%Fokker-Planck operator (\cite{coiffp}).

We choose the value of $\epsilon$ used in the DMAP
computation by invoking the \textit{correlation dimension} \cite{loglog}.
The assumption here is that the volume of an $n$-dimensional set
scales with any characteristic length $s$ as $s^n$; for relatively
uniform sampling one might expect the number of neighbors less than
$s$ apart to scale similarly.
We demonstrate this technique on a simple illustrative data set (points sampled
from a one-dimensional curve, see Figure \ref{fig:loglog}
below) by first computing all pairwise Euclidean distances.
Figure \ref{fig:loglog} shows $L(\epsilon)$, the total number of pairwise distances
less than $\epsilon$; it is clear that an asymptote will arise at
large $\epsilon$ ($M^2$, where $M$ is the number of points) and at
small $\epsilon$ ($M$).
In the figure, the two asymptotes are smoothly connected by an
approximately straight line;  the slope of this line suggests the
correct dimensionality for our data set (here, one).
The range of $\epsilon$ values corresponding to this straight
segment are all acceptable in our DMAP computations: here
any value of $\epsilon$ between approximately $10^{-2}$
and $10^{0}$ may be used.
Figure \ref{fig:loglog} also schematically illustrates the difference between PCA-based
and DMAP-based reduction; the data are in the form of a two-dimensional spiral embedded
in three-dimensional space.
Using the original coordinates to parametrize
this spiral requires three coordinates, while using PCA requires two and
DMAPs require only one (since the actual dimensionality of
the spiral is one).

\begin{figure}
\begin{center} \includegraphics[height=128mm,keepaspectratio=true,angle=270]{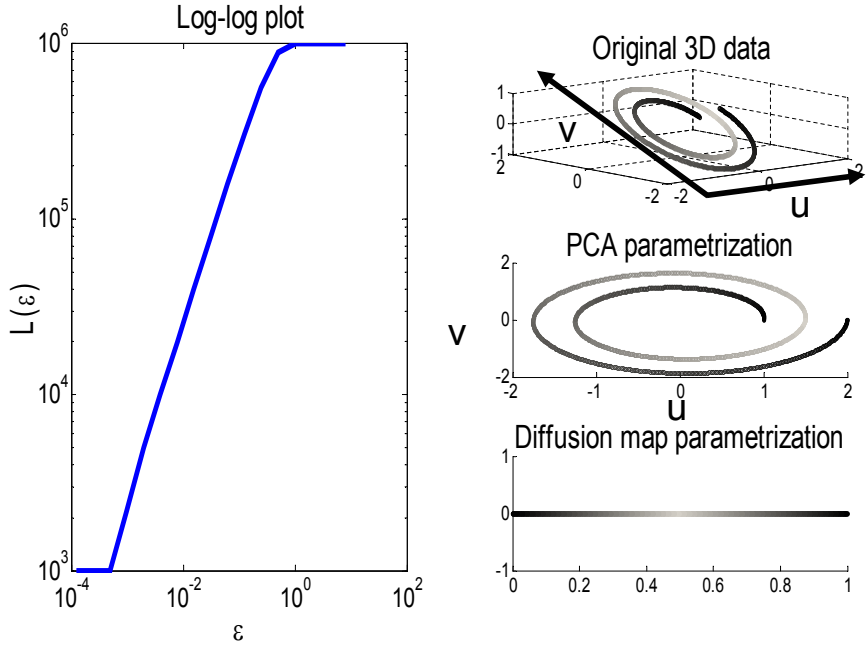}
    \caption{Left: log-log plot of $L(\epsilon)$, a statistic of
the data set (see text) vs. $\epsilon$.  For $M$ data vectors, $\lim_{\epsilon \to \infty} L(\epsilon)$ is
    simply $M^2$ and $\lim_{\epsilon \to 0} L(\epsilon)$ is
    simply $M$.  These two asymptotes are connected, however, by an approximately
    straight line.  $\epsilon$ values chosen from this regime are appropriate for our DMAP calculations.  
Right: three parametrizations of the same two-dimensional spiral
embedded in three-dimensional space (again, see text).}
    \label{fig:loglog}
\end{center} \end{figure}

\section{Ambient space to DMAP space and back}

In order to utilize the model reduction machinery provided by diffusion maps, one must be able to map back and forth between the original, high-dimensional ambient space and the reduced, low-dimensional DMAP space.  From the original space to DMAP space, there exists a mathematically elegant approach known as Nystr\"om extension; the reverse map, however, is more difficult.

\subsection{Nystr\"om extension}

The problem of finding the DMAP coordinates of a
\textit{new} $d$-dimensional vector $\vec X_{\mbox{new}}$ (a point not contained in the original
data set) is solved with the Nystr\"om
extension.
The first step is to compute all distances
$\{d_{i\,\text{new}}\}_{i=1}^M$ (or at least those which do not give
a negligible similarity measure) between our new vector and the $M$
vectors in our data set, and set, for instance $W(i,\text{new}) =
W(\text{new},i)= \text{exp} \left[ - \left(
\frac{d_{i\,{\text{new}}}}{\epsilon} \right)^2 \right]$ or
$W(i,\text{new}) =
W(\text{new},i)=(1-d_{i\,{\text{new}}}^2/\epsilon^2)
\mathbf{1}_{\{d_{i\,{\text{new}}}<\epsilon \}}$ (corresponding to
equations (\ref{eq:Gauss}) and (\ref{eq:Epan}), respectively).
Setting
\begin{equation*}
K(i,\text{new}) = \left(\sum_{k=1}^M W(k,\text{new})\right)^{-1}
W(i,\text{new}),
\end{equation*}
the $j\text{th}$ DMAP coordinate (there are $n$ coordinates total) of the new vector
$\vec X_{\mbox{new}}$ is given by
\begin{equation*}
\vec \psi_j(\vec X_{\mbox{new}}) = \frac{1}{\lambda_j} \sum_{i=1}^M K(i,\text{new})\, \vec \psi_j^{(i)}.
\end{equation*}
We typically drop the $n$ from $\vec \Psi_n$ (again, $n$ denotes the number of DMAP coordinates we choose to retain, and thus, the dimension of our DMAP space), and denote this process of Nystr\"om extension as $\vec \Psi(\vec X)$.
Clearly, this extension procedure will return a result even if $\vec X_{\mbox{new}}$ is not chosen to be \textit{exactly} on our low-dimensional manifold

\subsection{Construction of an ``inverse Nystr\"om map''} \label{invNystSec}

An important component of our reduced computations below is the ability to transform from DMAP space back to ambient space.
In other words, for given values of low-dimensional DMAP coordinates, we wish to find the corresponding ``on manifold'' point in the high-dimensional Euclidean space where the original data ensemble lies.
Approaches proposed include simulated annealing \cite{frewenchem}, Newton iteration, polynomial interpolation, 
interpolation using radial basis functions \cite{powell1987radial}, manifold regularization \cite{mreg}, and geometric harmonics \cite{gh}.
The method we chose here is polynomial interpolation of the ambient coordinates over DMAP space; each coordinate of ambient space is written as a polynomial over the low-dimensional DMAP space.
Polynomial interpolation is used simply because it is easy to derive order of magnitude error estimates; yet one should
mention (even though this was not observed in our computations) that singular matrices can, in principle, arise in the process (see, e.g. the discussion in \cite{powell1987radial})
The geometric harmonics extension scheme, akin to Fourier interpolation on manifolds, may very well outperform polynomial interpolation and is described below.

Since this transformation process is analogous to an \emph{approximate} inverse of the diffusion map (we go from low-dimensional DMAP space to high-dimensional ambient space), we denote it as $\vec \Psi_n^{-1}(\vec L)$, or simply $\vec \Psi^{-1}(\vec L)$, for $\vec L \in \mathbf{R}^n$ (DMAP space).

\subsubsection{Polynomial interpolation}
Suppose we wish to find $\vec X \in \mathbf{R}^d$ on the manifold such that for a given $\vec L \in \mathbf{R}^n$,
\begin{equation*}
\vec L = \vec \Psi_n(\vec X)\mbox{, or, as we indicated above, simply }\vec L = \vec \Psi(\vec X).
\end{equation*}
We first determine the $K$ closest points to $\vec L$ in our data
set (with proximity measured in DMAP space) and note both the DMAP
coordinates $\{ \vec L_l \}_{l=1}^K$ and corresponding ambient space
coordinates $\{ \vec X_l \}_{l=1}^K \in \mathbf{R}^d$ of these $K$
points.
These points are used to compute the coefficients of a local
polynomial interpolation of each ambient space coordinate over the DMAP coordinates; for each
coordinate $j=1,2,...,d$ of ambient space, we determine the
coefficients for the polynomial interpolation using $\{\vec L_1,\vec
X_1^{(j)}\},\{\vec L_2,\vec X_2^{(j)}\}, ...,\{ \vec L_K,\vec
X_K^{(j)}\}$, and we denote the result as $P^{(j)}$, setting
$\vec X^{(j)} = P^{(j)}(\vec L)$.
This procedure is relatively fast because the polynomials are
constructed over $\mathbf{R}^n$, and $n$ has been assumed small.
For example, in the reaction-diffusion example that follows, we
choose $K=10$ and interpolate 
over the {\em two-dimensional} DMAP space using six overall (up to quadratic) terms.
The six coefficients are fitted through least squares since, for $K=10$,
the system is overdetermined (only $K=6$ is required).
The result of this procedure is a point $\vec \Psi^{-1} (\vec L)$ which is an \textit{approximation} of the corresponding ``true,'' on-manifold point.
The error introduced by this approximation will affect our reduced dynamical model computations, as we will mention below.

\subsubsection{Geometric harmonics}
An additional computational tool from harmonic analysis which can be used for interpolation and
extrapolation is {\em geometric harmonics} \cite{gh,lafonthesis}.
We do not use this tool here, since it is difficult for us
to derive order-of-error bounds;  we do, however, feel that it is an important
option, and -while we omit many significant mathematical details-
we include here a brief description, for completeness.
The geometric harmonics extension scheme, inspired by the Nystr\"om
extension scheme, is a method for extending functions defined over
some smaller set $X$ to a larger one $\bar{X}$ \cite{gh}.  This is
exactly the problem of constructing the inverse Nystr\"om map; for
each point in our data ensemble, we know both its DMAP coordinates
and its ambient space coordinates (this can be viewed as $d$
functions from DMAP space ($\mathbf{R}^n$) to each coordinate in
ambient space), and we now wish to know this function for new values of the DMAP
coordinates as well. Throughout, we use $f$
to denote the function we desire to extend from $X$ to $\bar{X}$.

\begin{figure}
\begin{center} \includegraphics[height=45mm,keepaspectratio=true,angle=0]{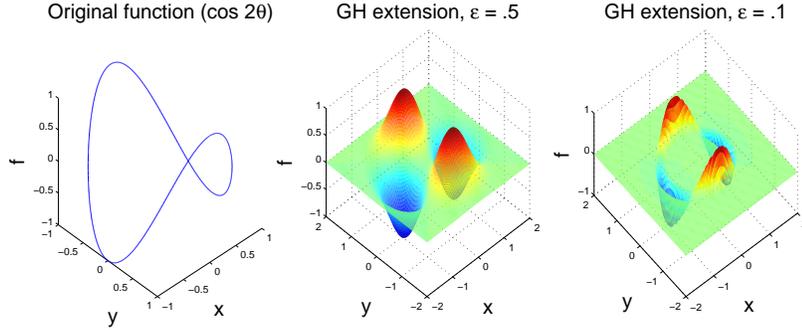}
    \caption{The original function of $f=\cos(2 \theta)$, defined on the unit circle (left), is extended with $\epsilon=0.5$ (middle) and $\epsilon=0.1$ (right).}
    \label{fig:GHext}
\end{center} \end{figure}

We first choose a symmetric, positive semi-definite kernel
$k:\,\bar{X}\times\bar{X}\rightarrow \mathbf{R}$ such as
$k(x,y)=\mbox{exp}\left(- \left\|x-y
\right\|^2/\epsilon^2\right)$.  Here, $\epsilon$ quantifies
the distance away from the data set $X$ that we wish to be able to
extend our function $f(\bar{X})$.  The key observation is that when
$f$ oscillates with frequency $\nu$, it cannot be extended reliably
beyond a distance of $O(1/\nu)$.  This makes intuitive sense, as
interpolation of functions $f$ which change rapidly should not be
trusted far from the ``known'' values of $f$.  For demonstration, a
geometric harmonics extension of the function $f(\theta) = \cos(2
\theta)$ is shown in Figure \ref{fig:GHext}, where $\theta =
\mbox{arctan} \frac{y}{x}$ and the set $X$ is simply the circle in
$\mathbf{R}^2$ given by $x^2+y^2=1$.  For $\epsilon=0.5$, the
extension of $f$ onto the plane is ``wider" (see Figure \ref{fig:GHext}) than for the case
$\epsilon=0.1$, but as we will see, this comes with the trade-off of
increased extension error (here, the error is measured on the
``known'' set $X$, see below).  Here, it is worth noting that, due to the nature of the kernel $k$ used, this formulation bears a strong resemblance to function approximation procedures using radial basis functions (RBFs) (see, e.g. \cite{fornberg_flyer}).

We can restrict this kernel to $X$ and define the linear operator $\mathbf{K}$ as
\begin{equation*}
 \mathbf{K}f(\bar{x}) = \int_{X} k(\bar{x},x)f(x)d\mu(x),
\end{equation*}
where $\mu(x)$ is just the measure of the set $X$ (if the data set
consists of $M$ random samples, $\mu(x) = 1/M$ is typically used).
It can be shown that this operator has a discrete set of eigenvalues
$\{\lambda_j\}\geq 0$ (in nonincreasing order) and orthonormal
eigenfunctions $\{\tau_j\}$ so that for almost all $x \in X$,
\begin{equation*}
 \lambda_j \tau_j(x) = \int_{X} k(x,y) \tau_j(y) d\mu(y).
\end{equation*}
Over finite sets, these integrals are obviously just sums and the eigenfunctions are just eigenvectors.  The geometric harmonics are defined as the extension of the eigenfunctions to $\bar{X}$ via
\begin{equation*}
 \tau_j(\bar{x}) = \frac{1}{\lambda_j} \int_X k(\bar{x},y)\tau_j(y)d\mu(y)
\end{equation*}
for $\lambda_j \neq 0$.  Since $\lambda_j \rightarrow 0$ as $j \rightarrow \infty$, this extension procedure becomes numerically ill-conditioned.  We pick a condition number\footnote{We note that, as discussed in \cite{fornberg_flyer}, ill-conditioning is not inherent to the problem, but rather, to the implementation.} $1/\delta$, and let $S_\delta = \{j \mbox{ such that } \lambda_j \geq \delta \lambda_0 \}$.  We can now extend $f(X)$ to $f(\bar{X})$ as follows:
\begin{remunerate}
 \item Project $f$ onto the numerically acceptable eigenfunctions via
\begin{equation} \label{eq:projf}
 f(x) \rightarrow \sum_{j \in S_\delta} \langle f,\tau_j \rangle_X \tau_j(x),\quad\mbox{where $\langle \, \rangle$ denotes inner product}.
\end{equation}
\item Use $\tau_j$ to extend $f$ on $\bar{X}$ as
\begin{equation*}
 f(\bar{x}) = \sum_{j \in S_\delta} \langle f,\tau_j \rangle_X \tau_j(\bar{x}).
\end{equation*}
\end{remunerate}

With a fine kernel (small $\epsilon$) there is negligible error in the projection of $f$ in equation (\ref{eq:projf}) (here, the error is measured on the ``known'' set $X$); the eigenvalues decay slowly enough so that most of the eigenfunctions can be used ($\lambda_j \geq \delta \lambda_0$ for most $j$).
However, $\tau_j(\bar{x})$ quickly decays to zero for any $\bar{x}$ much farther away from the data set $X$ than $\epsilon$ because there, the kernel function $k$ practically vanishes.
With a coarser kernel (large $\epsilon$), there may be more error in the projection projection of $f$ in equation (\ref{eq:projf}) (again measured on the ``known'' set $X$) since most eigenfunctions are neglected.
However, the domain of extendability is much larger.
Because of this, a multiscale extension scheme is usually used in practice.
In such a scheme, $f$ is first projected at a coarse scale (large $\epsilon$), and then the error in this initial coarse projection (the residual $f$) is projected at a finer scale (smaller $\epsilon$).
The error in this second, finer
projection is then projected at an even finer scale, and this process is continued until the total error shrinks below some desired threshold.
The sum of these projections yields the extended $f$ function.
Often, an initial $\epsilon = \epsilon_0$ is chosen, and then during projection $i$, $\epsilon = 2^{-(i-1)}\epsilon_0$ \cite{gh}.
$f$ is therefore extended as a linear combination of functions which oscillate with frequency $\nu_j$ and which vanish at a distance $O(1/\nu_j)$ from the set $X$.
Several results about the mathematical optimality of the extension of $f$ by this procedure have been obtained \cite{gh}.

\section{Dynamics in reduced spaces}

As we outlined in the introduction, we begin with the setting of a dissipative partial differential
equation of the form
\begin{equation*}
\frac{\partial}{\partial t}u + Au = F(u),
\end{equation*}
operating on a function $u(\vec x,t)$.
The usual stipulation is for $F$ to be globally Lipshitz continuous and at least $C^1$, while $A$ is a self-adjoint, compact linear operator; slightly different requirements are possible (see, e.g., \cite{foias1989exponential,aimshort,aim,garcia1998postprocessing,titi1990approximate}).
After a  Galerkin projection of this equation using the eigenfunctions of $A$,
one obtains a stiff system of ODEs which describes the evolution of the coefficients
of the projection of the solution (projections onto these eigenfunctions of $A$).
Although the spectrum of $A$ is typically countably infinite, one often truncates
the eigenfunction expansion with only $d$ eigenfunctions, where $d$ is chosen
based on some physical intuition or specified error.
Under appropriate conditions, one can also obtain
a low-dimensional manifold of dimension $n<d$ in phase space to which trajectories are
quickly attracted, due to the dissipative nature of the  PDE \cite{foias1989exponential,aimshort,aim,garcia1998postprocessing,titi1990approximate}.

To illustrate the application of model reduction to dissipative PDEs, we utilize the Chafee-Infante reaction-diffusion equation
\begin{equation} \label{eq:RD}
 u_t-\nu u_{xx}+u^3-u=0,
\end{equation}
with $\nu = 0.16$ and periodic boundary conditions $u(0,t)=u(\pi,t)=0$.  This equation is exactly of the form of equation (\ref{eq:AIM}) with $Au = \nu u_{xx}$ and $F(u) = u-u^3$, and it is known to have a two-dimensional inertial manifold \cite{jolly1989explicit}.

We first perform a Galerkin projection of this equation onto the first $d=10$ Fourier modes ($\sin x$, $\sin 2x$, $\ldots$, $\sin 10x$), resulting in an equation of the form of (\ref{mainEq}):
\begin{equation*}
 \frac{d \vec y}{dt} = \vec f(\vec y),
\end{equation*}
where $\vec y \in \mathbf{R}^{10}$ denotes the vector of coefficients of the Galerkin projection, $u(x,t) = \sum_{i=1}^{10} \vec y^{(i)}(t) \sin ix$.
In the literature, the $i$th Fourier mode is usually denoted as $a_i$ instead of $\vec y^{(i)}$; this will be the convention we adopt when referring to the Fourier modes of the reaction-diffusion equation.
Next, we obtain a data set $\vec X_1,\vec X_2,...,\vec X_M \in \mathbf{R}^{10}$ sampled from the two-dimensional, slow, inertial manifold by randomly generating initial conditions and integrating these initial conditions for a brief ($t=1$) amount of time (sufficient enough for a decay to the slow manifold).
Finally, we use these vectors $\vec X_1, \vec X_2,\ldots,\vec X_M$ in order to perform the model reduction outlined in \S \ref{sec:POD} and \ref{sec:DMAP} below.

%
%The low-dimensional manifold to which trajectories quickly decay will
%then be a subset of $d$-dimensional space.

\subsection{Proper orthogonal decomposition} \label{sec:POD}

Discretely sampling simulation trajectories after the
initial (brief) approach to the attracting manifold will give rise to an ensemble
of data points that have been computed and saved as vectors in $\mathbf{R}^d$.
We can attempt to compress these data by taking advantage of
the low dimensionality of the manifold that they live on; this is traditionally
accomplished through the use of Principal Component Analysis (PCA) on the data.
If we find that most of the energy (defined in terms of vector inner
products in $\mathbf{R}^d$) of the data can be spanned by the first $n$ eigenvectors
of the PCA decomposition, this suggests that the data can be efficiently described
as vectors in $\mathbf{R}^n$ (as the components of the data in the directions of the
first $n$ eigenvectors).
PCA, in addition to simple data compression, can also form the basis of
a systematic way to
\begin{remunerate}
 \item reduce the dynamics of the original high-dimensional (here, $d$-dimensional) dynamical system to $k$ dimensions
\item rewrite the correponding dynamics in the new, reduced, low-dimensional space
\end{remunerate}
using again a Galerkin procedure.
One starts with the data set $\vec X_1,\vec X_2,...,\vec X_M \in \mathbf{R}^d$
sampled from the (long term dynamics on the) low-dimensional, slow, invariant, attracting manifold.
These $\vec X$ may come from a variety of sources.  Here, we have chosen to focus on $\vec X$
which represent the coefficients of the spectral representation of solutions of equation ($\ref{eq:AIM}$),
but these $\vec X$ may also be the concentrations of various chemical species for the slow
manifold of a chemical kinetics problem.

POD finds (for each $n=1,2,...,d$) the projection onto the $n$th dimensional linear subspace which maximizes the variance of the projected data.
Let us denote the projection of a vector in $\vec X \in \mathbf{R}^d$ onto this linear subspace as $\mathbf{P}_n(\vec X)$.
Defining $\vec c^{(l)} = \frac{1}{M} \sum_{k=1}^M \vec X_k^{(l)}$, $\mathbf{P}_n(\vec X)$ can be written as $\mathbf{P}_n(\vec X)= \vec c+\sum_{k=1}^n \langle \vec X-\vec c,\vec p_k \rangle \vec p_k$.
Here, we have $\langle \vec p_k,\vec p_j \rangle = \delta_{kj}$, and the $\{\vec p_k \}_{k=1}^d$ are the eigenvectors of the matrix $\mathbf{\tilde X} \mathbf{\tilde X}^T$ with
\begin{equation*}
\mathbf{\tilde X} = \left[\begin{array}{cccc} | & | & & | \\ \vec X_1-\vec c & \vec X_2-\vec c & ... & \vec X_M-\vec c \\ | & | & & | \end{array} \right].
\end{equation*}
It turns out that regardless of the final dimension $n$ of the linear subspace, the first projection direction remains optimal, as does the second, third, etc.; these $\{\vec p_k\}_{k=1}^n$ do not change with $n$.
Defining $\vec L \in \mathbf{R}^n$ as the coefficients of the projection in this reduced space, $\vec L^{(1)} = \langle \vec X-\vec c,\vec p_1 \rangle$, $\vec L^{(2)} = \langle \vec X-\vec c,\vec p_2 \rangle$, etc., we write $\vec L = \vec P_n(\vec X)$.
The Galerkin reformulation then yields the reduced dynamics of the $\vec y$ of equation (\ref{mainEq}) in this $n$-dimensional space.  If $\vec L = \vec P_n(\vec y)$, then:
\begin{eqnarray*}
\frac{d \vec L}{dt}  &=& \frac{d}{dt} \vec P_n(\vec y) \notag \\
&=& \vec P_n \left(\frac{d}{dt} \vec y \right) \notag
\end{eqnarray*}
because $\vec P_n$ is linear.  Continuing,
\begin{eqnarray*}
\frac{d \vec L}{dt} &=& \vec P_n\left( \vec f(\vec y) \right) \notag \\
&\approx& \vec P_n\left( \vec f\left(\vec c+\sum_{k=1}^n \vec L^{(k)} \vec p_k\right) \right)
\end{eqnarray*}
so that now things are expressed in terms of $\vec L$.
Evaluating the right hand side is a challenge often requiring quadrature or involving cumbersome analytical formulas
(unless $f$ is linear).

\begin{figure}
 \begin{center} \includegraphics[height=90mm,keepaspectratio=true,angle=0]{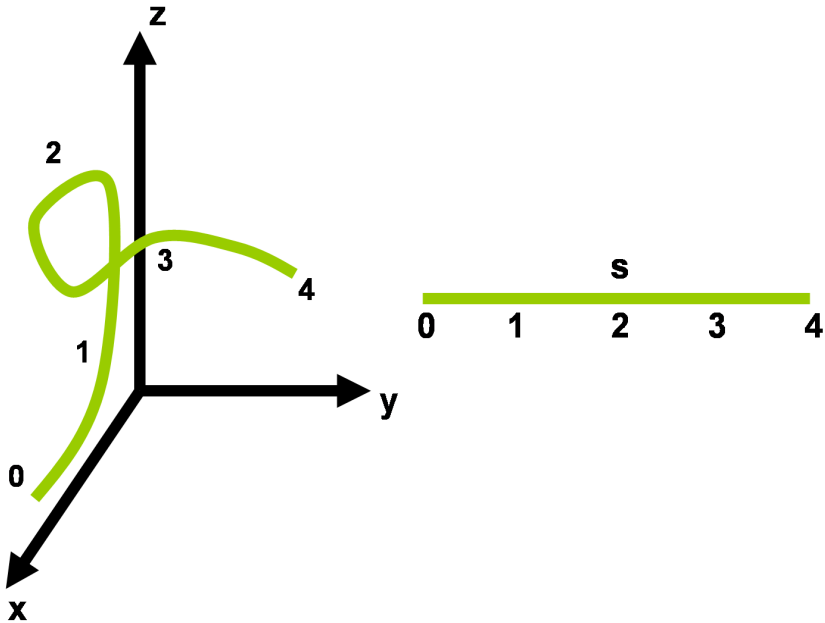}
    \caption{A one-dimensional curve, representing a one-dimensional slow manifold in three-dimensional space, is shown (left).  The curve is not a straight line, and it requires three POD modes to be well represented.  Ideally, we would like to write the slow dynamics in terms of one ``mode;'' for example, $(x(s),y(s),z(s))$ as a function of arclength $s$ (right).  Such a parametrization can be found using DMAPs.}
    \label{fig:modes}
\end{center} \end{figure}

\begin{figure}
 \begin{center} \includegraphics[height=45mm,keepaspectratio=true,angle=0]{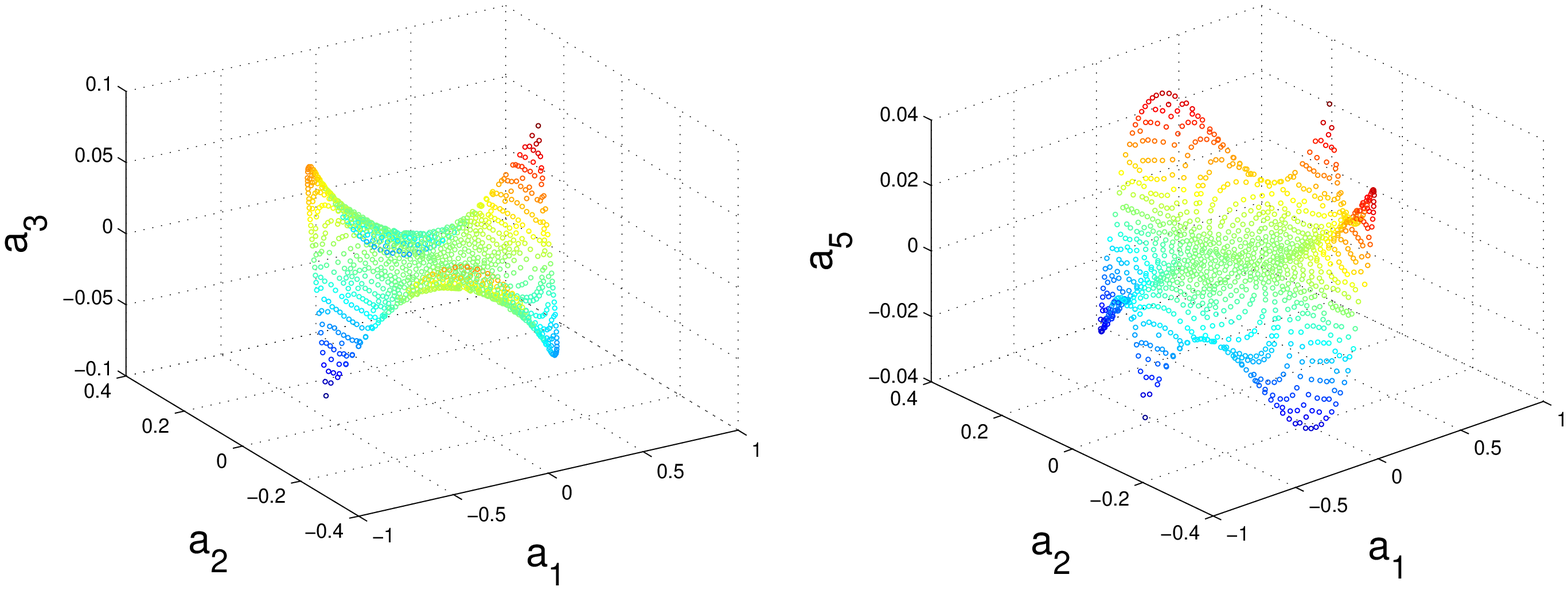}
    \caption{For a particular class of reaction-diffusion equations and for certain parameter values, an inertial manifold of dimension $2$ exists \cite{jolly1989explicit}, parametrized in Fourier space by the first two Fourier modes $(a_1,a_2)$.  For visualization purposes, we have shown two ``slices'' of this inertial manifold, $a_3(a_1,a_2)$ and $a_5(a_1,a_2)$,  in three-dimensions (with points colored by $a_3$ and $a_5$, respectively).  It is clear that both of these slices are highly nonplanar and that any sort of two-dimensional POD plane would do a poor job passing through or near all these points.}
    \label{fig:RD_manifold}
\end{center} \end{figure}

When the slow manifold of the system is not a linear subspace, this method
will not give much information
about the underlying dimensionality of the slow manifold because the number of modes used is
often based on some sort of error control and not on manifold geometry.
It will also take more
modes (larger $n$) than the dimensionality of the slow manifold to accurately represent the dynamics.
Figure \ref{fig:modes} illustrates this by showing a one-dimensional (nonlinear) curve lying in three-dimensional space, respresentative
of the possibility of one-dimensional dynamics in a higher-dimensional space.
Figure \ref{fig:RD_manifold} shows, in terms of a Fourier basis for the solution (again, denoted $a_1,a_2,\ldots$) of the reaction-diffusion equation (\ref{eq:RD}), two slices of the two-dimensional (but nonlinear) inertial manifold of the (high-dimensional) reaction-diffusion equation; again, the long-term dynamics are low-dimensional while
the ambient space dimension is high.
In these situations, a technique is desired in which we may
write the reduced dynamics of equation (\ref{mainEq}) on the actual, low-dimensional
slow manifold regardless of its shape or of whether it can be approximated by a hyperplane.

%Using a string of local PCAs instead of one global PCA \textit{will} provide information about the dimensionality of the slow manifold and allow %us to remain fairly close to the manifold to maintain accuracy (as long as each coordinate patch is sufficiently small), but then we are forced to %do multiple dimensionality reductions, glue together coordinate patches, etc.

%\begin{figure}
% \includegraphics[height=70mm,keepaspectratio=true,angle=0]{RD_PCA.eps}
%    \caption{For a particular reaction-diffusion equation, there is an inertial manifold of dimension $2$ in Fourier space (rainbow manifold).  %However, trying to ``span'' this inertial manifold using only $2$ principal components fails (blue plane).}
%    \label{fig:RD_PCA}
%\end{figure}

\subsection{The DMAP approach} \label{sec:DMAP}

In contrast to POD, whose model reduction is based upon the discovery of a linear subspace which {\em approximately} contains the slow manifold, diffusion maps allow us to write the dynamics of equation (\ref{mainEq}) on the ``true'' low-dimensional (and possibly nonlinear) slow, attractive, invariant manifold.

In this section, we use DMAPs to find a parametrization of the slow manifold and denote the dimensionality reduction procedure as $\vec L = \vec \Psi_n(\vec X)$, or, as before, we drop the $n$ and write $\vec L = \vec \Psi(\vec X)$, where $\vec L \in \mathbf{R}^n$.
Here, $\Psi$ is a transformation from the high-dimensional ambient space to the low-dimensional parametrization.
In our working example, the Chafee-Infante reaction-diffusion equation (\ref{eq:RD}), the inertial manifold is two-dimensional; we therefore expect $\vec \Psi_2$ to give us our manifold parametrization.
In Figure \ref{fig:RD_manifold}, we show a picture of the two-dimensional reaction-diffusion inertial manifold in only three-dimensional ambient space for visualization purposes.%  In actuality, $d=10$ for this illustrative dynamical system.
%The parametrization of the manifold we obtain after the dimensionality reduction by DMAPs is shown in figure(\ref{fig:2dMan}).  The %two-dimensional structure of the ten-dimensional data set is discovered.
%\begin{figure}
% \includegraphics[height=70mm,keepaspectratio=true,angle=0]{2dMan.eps}
%    \caption{The manifold ``learned'' by the DMAP from data points sampled from the inertial manifold of the reaction-diffusion equation %is, in fact, two dimensional (this can be seen by looking at the eigenvalue spectrum of the matrix $K$ from above).  Its parametrization by the %DMAP is shown here.  The first two coordinates $\vec \Psi_2^{(1)}$ and $\vec \Psi_2^{(2)}$ of the DMAP $\vec \Psi_2$.}
%    \label{fig:2dMan}
%\end{figure}

We now derive the reduced dynamics of equation (\ref{mainEq}).  If $\vec L = \vec \Psi(\vec y)$ (and
$\vec y = \vec \Psi^{-1}(\vec L)$), then:
\begin{eqnarray*}
\frac{d \vec L}{dt}  &=& \frac{d}{dt} \vec \Psi(\vec y) \\
&=& \frac{\partial \,\vec \Psi(\vec y)}{\partial \vec y}\,\frac{d \vec y}{dt} \notag
\end{eqnarray*}
by the chain rule, where $\frac{\partial \,\vec \Psi(\vec y)}{\partial \vec y}$ is a $n \times d$ matrix and $\frac{d \vec y}{dt}$ is a $d$-dimensional vector.
Continuing, 
\begin{eqnarray} \label{nonLinGalDyn}
\frac{d \vec L}{dt} &=& \frac{\partial \,\vec \Psi(\vec y)}{\partial \vec y}\,\vec f(\vec y) \notag \\
&=& \frac{\partial \,\vec \Psi \left(\vec \Psi^{-1}\left(\vec L \right) \right)}{\partial \vec y} \,\vec f\left(\vec \Psi^{-1}\left(\vec L \right) \right) 
\end{eqnarray}
so that now things are expressed in terms of $\vec L$.
Ways to obtain $\Psi^{-1}\left(\vec L \right)$ were discussed above (we used polynomial interpolation).
It is worth noting again that Nystr\"om extension, $\vec \Psi(\vec y)$, will also return values for points $\vec y \in \mathbf{R}^d$ which are neither \textit{exactly} on our manifold nor in the original data ensemble. 

When new points $\vec X_{\mbox{new}}$ (points not in our original data ensemble) such as $\vec \Psi^{-1}
(\vec L ) $ arise in the computation, we obtain $\frac{\partial
\,\vec \Psi}{\partial \vec y}$ at such points as follows:
\begin{eqnarray*}
 \vec \Psi^{(j)}(\vec X_{\mbox{new}}) &=& \frac{1}{\lambda_j} \sum_{i=1}^M K(i,\text{new})\, \vec \psi_j^{(i)} \notag \\
 &=& \frac{1}{\lambda_j} \frac{\sum_{i=1}^M W(i,\text{new})\, \vec \psi_j^{(i)}}{\sum_{i=1}^M W(i,\text{new})}. \notag
\end{eqnarray*}
Therefore,
\begin{eqnarray} \label{nastyDeriv}
 \frac{\partial \vec \Psi^{(j)}}{\partial \vec X_{\mbox{new}}^{(k)}}&=& \frac{1}{\lambda_j}\, \frac{\partial}{\partial \vec X_{\mbox{new}}^{(k)}} \,\left( \frac{\sum_{i=1}^M W(i,\text{new})\, \vec \psi_j^{(i)}}{\sum_{i=1}^M W(i,\text{new})} \right) \notag \\
%&=& \frac{1}{\lambda_j} \, \frac{\left(\sum_{i=1}^M  \frac{d}{d \vec X_{\mbox{new}}^{(k)}} W(i,\text{new})\, \vec %\psi_j^{(i)}\right)\left(\sum_{l=1}^M W(l,\text{new}) \right)}{\left(\sum_{i=1}^M W(i,\text{new})\right)^2} \notag \\
%& & -\frac{1}{\lambda_j} \,\frac{\left(\sum_{l=1}^M W(l,\text{new}) \, \vec \psi_j^{(l)}\right) \left(\sum_{i=1}^M  \frac{d}{d \vec %X_{\mbox{new}}^{(k)}} W(i,\text{new}) \right)}{\left(\sum_{i=1}^M W(i,\text{new})\right)^2} \notag \\
&=& \frac{1}{\lambda_j} \frac{\sum_{l,i=1}^M \, W(l,\text{new})
\left(\frac{\partial}{\partial \vec X_{\mbox{new}}^{(k)}}
W(i,\text{new}) \right)\left(\vec \Psi_j^{(i)} - \vec \Psi_j^{(l)}
\right)}{\left(\sum_{i=1}^M W(i,\text{new})\right)^2} .
\end{eqnarray}
It is really only necessary to compute $W(i,\text{new})$ for those $\vec X$ which lie near $\vec X_{\mbox{new}}$ since for distances beyond $\epsilon$, contributions from other points either dies off (Gaussian kernel) or disappears completely (Epanechnikov kernel).

Now that we have these formulas, we can compute the dynamics of equation (\ref{nonLinGalDyn}) starting from an initial condition in DMAP space.
Even if this initial condition is selected to be one of the known data points, immediately after the first integration step, we will find the trajectory visiting ``new'' points $\vec X_{\mbox{new}}$ that do not belong to our original data set.
When such new points arise, we are able to compute the necessary time derivatives via equation (\ref{nastyDeriv}).
Armed with these formulas, we can now continue the integration of equation (\ref{nonLinGalDyn}), constantly going back and forth between DMAP and ambient representations in order to obtain the time derivatives in DMAP space.

\begin{figure}
 \begin{center} \includegraphics[height=70mm,keepaspectratio=true,angle=0]{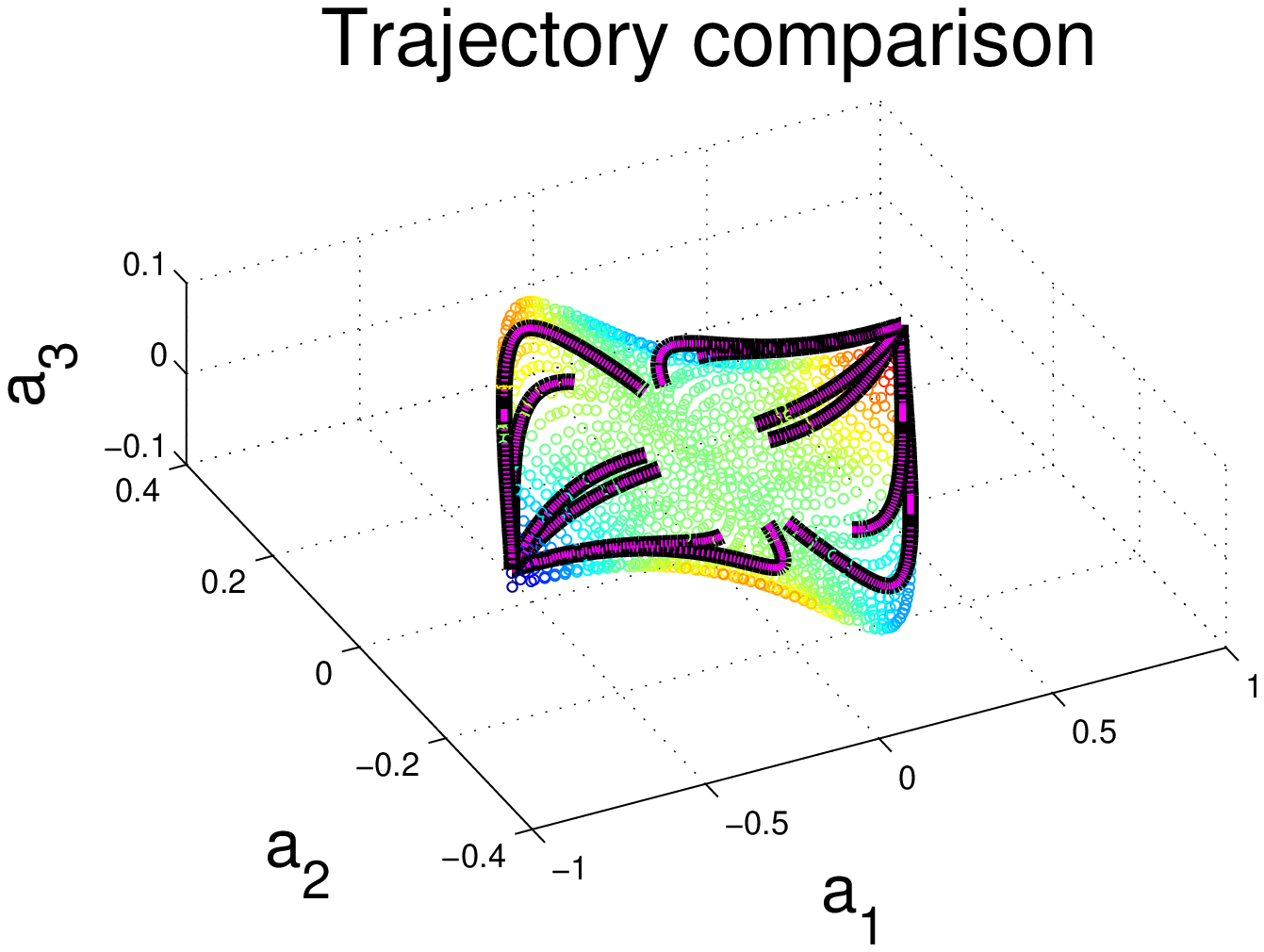}
    \caption{A visual comparison of the trajectories obtained through the original dynamics of equation (\ref{mainEq}) (black) and the new dynamics of equation (\ref{nonLinGalDyn}) (purple), shown for visualization purposes in three-dimensional space (points are colored by their value of $a_3$).  To perform this comparison, the trajectory corresponding to the reformulated dynamics had to be transformed into the original space using the $\vec \Psi^{-1}(\vec L)$.}
    \label{fig:TrajComp}
\end{center} \end{figure}

To validate our approach, we compare two finite trajectories:
\begin{remunerate}
\item a trajectory begun at $\vec X(0) \in \mathbf{R}^d$, integrated according to equation (\ref{mainEq}) for time $t$
 \item a trajectory begun at $\vec \Psi(\vec X(0)) \in \mathbf{R}^n$ (DMAP space), integrated according to equation (\ref{nonLinGalDyn}) for time $t$, and transformed back into the original, ambient space of $\mathbf{R}^d$ using $\vec \Psi^{-1}$.
\end{remunerate}
We integrated these two trajectories very accurately using ode113 and absolute and relative error tolerances of 1e-7.
A particular slice of these two $10$-dimensional trajectories is shown in Figure \ref{fig:TrajComp}; clearly, they visually coincide (and do so for all other slices as well).

It is also interesting to observe that reduction in the number of variables
also reduces the model problem's stiffness.
Numerically, we observe that for any $d$, the magnitude of the
largest absolute eigenvalue of the Jacobian of the reformulated system
(\ref{nonLinGalDyn}) for our problem is $O(1)$, whereas the magnitude
of the maximum absolute eigenvalue of the Jacobian of the original system
(\ref{mainEq}) is $O(d^2)$.
The reason is that derivatives of
$\vec \Psi(\vec X)$ are zero in directions orthogonal to the
slow manifold, so stiff directions (which in this example are
approximately orthogonal to the manifold) are effectively projected
out of equation (\ref{nonLinGalDyn}).
In fact, for certain error tolerances, computing the trajectories
shown in Figure \ref{fig:TrajComp} using our approach takes less
wall clock time than computing the trajectories via the original dynamics
(\ref{mainEq}).
With absolute and relative error tolerances of $10^{-6}$, for
instance, ode45 spends $2.11$ seconds of wall clock time to compute
the ``new" dynamics, while the original dynamics takes $5.73$ seconds
and three times as many calls to the derivative function.
Explicit integrators are less susceptible to stability constraints
due to the reduced stiffness (allowing fewer time steps), while
implicit integrators fare well for an entirely different reason:
the reduced dimensionality of the problem makes the computational
linear algebra (in the Newton iterations required to solve the
corresponding nonlinear equations of the implicit integrator)
faster.
Our purpose in this paper is to implement and discuss the procedures
necessary for the formulation and solution of the reduced
equations (\ref{nonLinGalDyn}); a more quantitative comparison of the
cost of integration of equations (\ref{mainEq}) and (\ref{nonLinGalDyn})
is the subject of current research.

%\subsection{Some accuracy considerations}

There are two reasonable ways of quantifying the difference between
the two sets of dynamics, the original dynamics (\ref{mainEq}) and the ``new'' dynamics of our approach (\ref{nonLinGalDyn}).
The first is to compare,
in ambient space, the trajectories computed by our approach with those computed according to the original
dynamics as in Figure \ref{fig:TrajComp}.
The second is to compare these two trajectories in DMAP space instead.
Comparing trajectories in either DMAP space or ambient space
requires first obtaining these trajectories, and thus
making choices about the particular integrators; however, even when the two sets of equations (the original and the new)
are integrated with great numerical accuracy, independent of cost,
their trajectories will differ because of the interpolation (the $\vec \Psi^{-1}$ map) in equation (\ref{nonLinGalDyn}).
Furthermore, the original dynamics have a different dimensionality and stiffness than the dynamics of our approach.
We therefore choose to focus on only the error involved in the 
evaluation of the {\em right hand side}
of equation (\ref{nonLinGalDyn}), the error in the DMAP-based dynamics.

Suppose the integrator (in DMAP space) produces a new point $\vec L$ (in DMAP space).
The derivative at $\vec L$ constructued by equation (\ref{nonLinGalDyn}) is only approximate because the $\vec \Psi^{-1}$ map is not exact (and this equation uses the $\vec \Psi^{-1}$ map twice).
We use $\Delta \vec x$ to denote the $\vec \Psi^{-1}$ interpolation error\footnote{It is possible for $\Delta \vec x$ (the error in $\vec \Psi^{-1}$) to become large near the boundary of the manifold in DMAP space \cite{lafonthesis} unless the DMAP algorithm is slightly modified; although we do not observe this behavior (Figure \ref{fig:manErrComp}, for instance, shows the $\vec \Psi^{-1}$ error to be negligible), we note it and avoid integrating trajectories in this region.}:
\begin{equation*}
 \vec \Psi^{-1}\left(\vec L\right) = \vec \Psi^{-1}_{\mathbf{true}}\left(\vec L\right) + \Delta \vec x.
\end{equation*}
Now, we wish to study the error in equation (\ref{nonLinGalDyn}) due to $\Delta \vec x$.
Restricing our attention to a single coordinate
$g \in 1,2,...,n$ of DMAP space, equation (\ref{nonLinGalDyn}) takes the following form:
\begin{eqnarray} \label{eq:error}
\frac{d\vec L^{(g)}}{dt} &=& \frac{\partial \,\vec \Psi^{(g)} \left(\vec \Psi^{-1}\left(\vec L \right) \right)}{\partial \vec y} \,\vec f\left(\vec \Psi^{-1}\left(\vec L \right) \right) \notag \\
&=& \frac{\partial \,\vec \Psi^{(g)} \left(\vec \Psi_{\mathbf{true}}^{-1}\left(\vec L \right)+\Delta \vec x \right)}{\partial \vec y} \,\vec f\left(\vec \Psi_{\mathbf{true}}^{-1}\left(\vec L \right)+\Delta \vec x \right),
\end{eqnarray}
where, now, $\frac{\partial \,\vec \Psi^{(g)}}{\partial \vec y}$ is a $1 \times d$ matrix.
Expanding equation (\ref{eq:error}), and keeping only first order terms, we see that the error in coordinate $g$ is given as:
\begin{eqnarray} \label{eq:errExp}
\mbox{err}^{(g)} &\equiv& \frac{d \vec L^{(g)}}{dt}-\frac{\partial \,\vec \Psi^{(g)} \left(\vec \Psi_{\mathbf{true}}^{-1}\left(\vec L \right) \right)}{\partial \vec y} \,\vec f\left(\vec \Psi_{\mathbf{true}}^{-1}\left(\vec L \right) \right) \notag \\
&\approx& \Delta \vec x^T \,\,\frac{\partial ^2\,\vec \Psi^{(g)}\left(\vec \Psi_{\mathbf{true}}^{-1}\left(\vec L \right)\right)}{\partial  \vec y^2} \vec f \left(\vec \Psi_{\mathbf{true}}^{-1}\left(\vec L \right)\right) \\
& &\quad + \frac{\partial \,\vec \Psi^{(g)}}{\partial \vec y}\left(\vec \Psi_{\mathbf{true}}^{-1}\left(\vec L \right)\right) \mathbf{J}\left(\vec \Psi_{\mathbf{true}}^{-1}\left(\vec L \right)\right) \Delta \vec x, \notag
\end{eqnarray}
where $\frac{\partial ^2\,\vec \Psi^{(g)}}{\partial  \vec y^2}$ and $\mathbf{J}$ are $d \times d$ matrices (and $\mathbf{J}$ is the Jacobian of the function $\vec f$).
Combining these terms, we see that the overall error in the $g$th component is bounded by
\begin{equation*}
 \left|\mbox{err}^{(g)}\right| \leq \left|\left|\Delta \vec x \right| \right| \left(\left|\left|\frac{\partial ^2\,\vec \Psi^{(g)}}{\partial  \vec y^2}\right|\right|\left|\left| \vec f \, \right| \right| + 
\left|\left|\frac{\partial \vec \Psi^{(g)}}{\partial \vec y} \mathbf{J} \right|\right| \right),
\end{equation*}
where $\|\,\,\|$ represents the matrix (vector) norm.
Here, $\left\| \vec f \, \right\|$ should be small since it is evaluated ``on manifold'' at point $\vec \Psi_{\mathbf{true}}^{-1}\left(\vec L \right)$; essentially, since we are on the slow manifold, only slow derivatives are present.
$\left\|\frac{\partial^2\,\vec \Psi^{(g)}}{\partial \vec y^2}\right\|$ can be shown to be bounded and $O(1)$ under appropriate scaling \cite{lafonthesis}.
Finally, when the fast directions (eigenvectors of $\mathbf{J}$ with large negative eigenvalues) are approximately
orthogonal to the manifold, as they are in our reaction-diffusion illustrative example,
the norm of the $n \times d$ matrix $\frac{\partial \,\vec \Psi}{\partial \vec y} \mathbf{J}$ will be small (on the order of the eigenvalues of the slow subspace); the kernel of $\frac{\partial \,\vec \Psi}{\partial \vec y}$ consists of the directions orthogonal to the manifold.

%
%We note that the term in parentheses is small since $\left|\left| \vec f \, \right| \right|$ and $\left|\left|\frac{\partial \vec \Psi^{(g)}}{\partial %\vec y} \mathbf{J} \right|\right|$ are small as well (on the order of slow manifold derivatives and eigenvalues, respectively).

\subsection{Nonlinear Galerkin techniques} \label{sec:NonLinGal}
An alternative approach to obtaining accurate reduced discretizations of
dissipative PDEs is provided by the so-called ``nonlinear Galerkin'' techniques in the
context of approximate inertial manifolds.
Instead of a Galerkin projection onto a large number of orthogonal eigenfunctions,
leading to a large set of coupled ODEs, one (approximately) expresses the component of the solution
in the ``higher" modes as a function of its components in the ``lower" modes
(see, e.g., \cite{initmanbook,constantin1985attractors,foias1988inertial,foias1989exponential,aimshort,aim,NLGreview,garcia1998postprocessing,titi1990approximate}).

For the two-dimensional reaction-diffusion inertial manifold above, for
instance, it is known that one can approximate the coefficients of each of the higher-order Fourier modes ($a_3,a_4,a_5,...$)
on the slow manifold as functions of just the first two Fourier modes, $a_1$ and $a_2$.
The nonlinear Galerkin formulation will then depend on only two variables, the
first two Fourier modes, and the reduced dynamics are two-dimensional.
%
%To use a nonlinear Galerkin method, it must be either known or assumed that high-order modes are
%slaved to low-order modes, and a way of retrieving the functional form of these relationships must
%exist.
Two popular techniques to approximate the slaving of the higher to the lower-order modes
include the so-called ``steady'' and  the ``Euler-Galerkin'' approximate intertial manifolds (AIMs).

The steady approximation for our reaction-diffusion problem is implemented by setting $\dot a_3 = 0$,
$\dot a_4 = 0$, $\dot a_5 = 0$, ..., from which we obtain the functions $a_3(a_1,a_2)$, $a_4(a_1,a_2)$,
$a_5(a_1,a_2)$, ...
%
%For instance, setting $\dot a_3 = 0$, we obtain
%\begin{equation}
%a_3(1-9 \nu - 6 a_2^2-6 a_1^2)-3 a_3^3 = 3 a_1 a_2^2 - a_1^3,
%\end{equation}
%which can be solved
%with a Newton iteration ($\nu$ is just a parameter for the reaction-diffusion equation).
%
We evolved the steady approximate inertial form using MATLAB's built-in Differential Algebraic Equation (DAE) solver ode15s.
At 3.44 seconds, this code took longer than both the original and the DMAP-based dynamics; this may very well be because, even though the system is conceptually two-dimensional, the linear algebra computations in the DAE solver are performed in a higher (here, $10$) dimensional space.

In the Euler-Galerkin approximation the functions $\{a_k(a_1,a_2)\}_{k}$
are obtained by constraining $a_1$ and $a_2$ to be constant, and taking an
implicit Euler step (for appropriately chosen size $\tau$) of the constrained
dynamics of the higher-order Fourier modes.
Here, as in \cite{aimshort}, we used $\tau=1$, which is large enough for the higher-order modes to get slaved to the first
two modes.
Instead of solving the nonlinear equations resulting from the implicit Euler step, a
fixed point iteration is implemented (and in fact, only a single substitution
is performed, since the map is so contracting).
For simplicity, we consider only $a_1,a_2,$ and $a_3$ and set $a_4 = a_5 = \cdots = 0$.
Then the constrained $a_3$ dynamics are given by
\begin{eqnarray*}
\dot a_3 &\equiv& \dot a_{3\,\mbox{diffusive}} + \dot a_{3\,\mbox{reactive}} \\
&=& (-9\nu a_3)+(a_3-3a_3^3-6a_2^2a_3-6a_1^2a_3+a_1^3-3a_1a_2^2). \notag
\end{eqnarray*}
Taking one implicit Euler step of length $\tau$ while holding $a_1$ and $a_2$ constant, and using  as initial
condition $a_3(0) = 0$ ($0$ is a good initial guess since higher-order modes quickly become small), we obtain
\begin{eqnarray*}
 a_3(\tau) &=& a_3(0)+\tau \dot a_3(\tau) \\
&=& 0+\tau \left[\dot a_{3\,\mbox{diffusive}}(a_1,a_2,a_3(\tau)) + \dot a_{3\,\mbox{reactive}}(a_1,a_2,a_3(\tau)) \right]. \notag
\end{eqnarray*}
Moving the diffusive term to the left hand side,
\begin{equation} \label{eq:EG}
(1+9 \tau \nu) a_3(\tau) = \tau[a_3(\tau)-3a_3^3(\tau)-6a_2^2a_3(\tau)-6a_1^2a_3(\tau)+a_1^3-3a_1a_2^2].
\end{equation}
The map $a_3 \leftarrow \frac{\tau}{1+9\tau\nu}(a_3-3a_3^3-6a_2^2a_3-6a_1^2a_3+a_1^3-3a_1a_2^2)$ can be
shown to be a contraction, intuitively because of the large eigenvalues associated with the diffusion
term.
Hence, instead of solving equation (\ref{eq:EG}), we invoke the method of successive substitutions
with initial guess $a_3 = 0$ to obtain the approximation $a_3 = \frac{\tau}{1+9\tau\nu}(a_1^3-3a_1a_2^2)$.

\begin{figure}
 \begin{center} \includegraphics[height=48mm,keepaspectratio=true,angle=0]{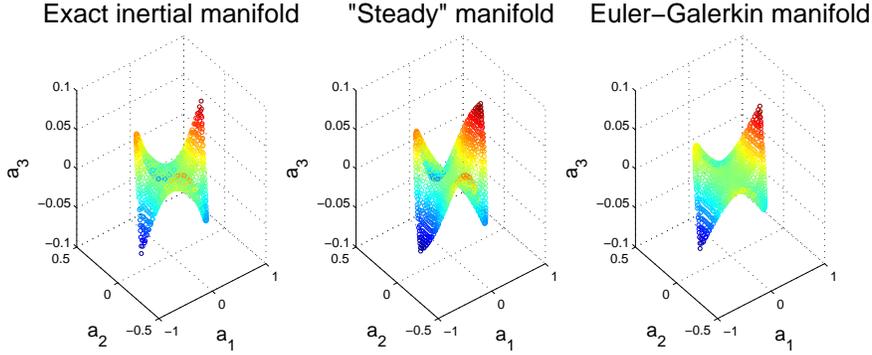}
    \caption{A comparison of the exact inertial manifold with the two nonlinear Galerkin AIMs
described in \S \ref{sec:NonLinGal} (see text).}
    \label{fig:manComp}
\end{center} \end{figure}

\begin{figure}
 \begin{center} \includegraphics[height=50mm,keepaspectratio=true,angle=0]{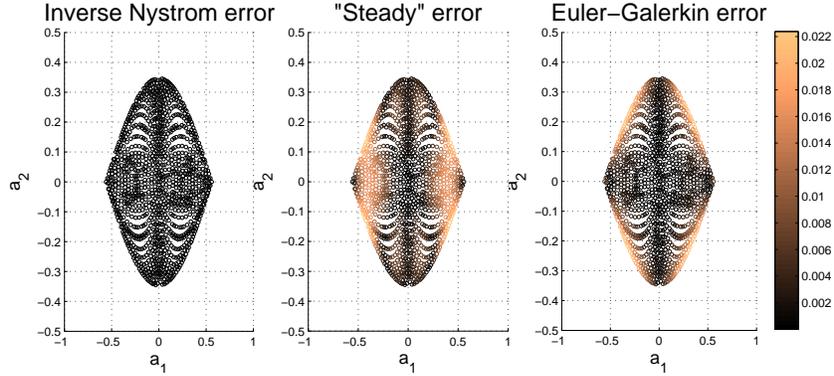}
    \caption{A comparison of the errors of our inverse Nystr\"om manifold (the ``interpolated manifold''
used to evolve equation (\ref{nonLinGalDyn})), the steady AIM, and the
Euler-Galerkin AIM (see text).}
    \label{fig:manErrComp}
\end{center} \end{figure}

A comparison of the exact manifold (from accurate full simulations), the steady AIM, and the Euler-Galerkin AIM can be seen in Figure
\ref{fig:manComp}.
In Figure \ref{fig:manErrComp}, we show the magnitude of the error (the norm of
the vector
\begin{equation*}
\left[(a_{3\,\mbox{computed}}-a_{3\,\mbox{exact}})\quad (a_{4\,\mbox{computed}}-a_{4\,\mbox{exact}})\quad ... \quad \right]
\end{equation*}
of errors) for the manifold constructed by our inverse Nystr\"om map (essentially a polynomial interpolation
of points on the exact manifold), the steady AIM, and the Euler-Galerkin AIM.

\begin{figure}
\begin{center}
\includegraphics[height=50mm,keepaspectratio=true,angle=0]{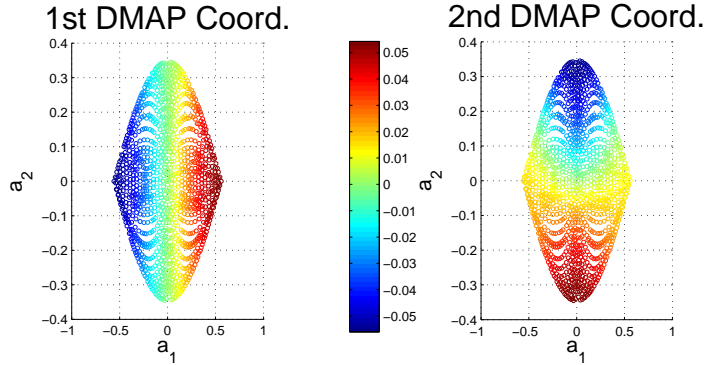}
    \caption{A plot of the first DMAP coordinate (left) and the second DMAP coordinate (right) vs. the first two Fourier modes ($a_1$ and $a_2$).  The figure strongly suggests that the determinant of the Jacobian is everywhere nonzero, or equivalently, that the transformation between the first two DMAP coordinates and the first two Fourier modes is unique;  on the left, $a_1$ appears to be one-to-one with the first DMAP coordinate, and on the right, once a value of $a_1$ (equivalently, the first DMAP coordinate) is specified, $a_2$ one-to-one with the second DMAP coordinate.}
    \label{fig:DMvsAPlot}
\end{center} \end{figure}

We verified numerically that an isomorphism exists between the first two DMAP
coordinates and the first two Fourier modes for points on the inertial manifold.
This was established by verifying that the determinant of the Jacobian
of the transformation from two-dimensional DMAP space to $(a_1, a_2)$ Fourier space is everywhere nonzero.
This nonsingularity of the Jacobian is also suggested by Figure \ref{fig:DMvsAPlot}.
As
expected, the DMAP correctly captures the two-dimensional geometry of the slow manifold;
the DMAP eigenvalues we computed for the data using $\epsilon=0.2$ are
$(1.0,0.8477,0.7504,0.4585,0.4155,...)$, and
the first two nontrivial eigenvalues are clearly larger than the rest.

\section{Conclusion}
In this paper we demonstrated a link between manifold learning techniques (and, in particular,
diffusion maps) and model reduction for dissipative evolutionary PDEs.
The approach is data-based, and it provides an interesting (``nonlinear") alternative to the the well-known Galerkin
projection of the dynamics on the leading principal components of the data ensemble.
The implementation of the algorithm is only slightly more involved than that of the classic POD-Galerkin
method, and is often faster due to decreased stiffness and a more parsimonious dimensionality reduction;
the latter case occuring with low-dimensional but highly nonlinear (non-flat)
slow manifolds residing in high-dimensional ambient spaces.
%
%At the same time, the resulting dynamics are more representative of the true reduced
%behavior of the system.

Instead of evaluating $d \vec L/dt$ from the right hand side of equation (\ref{nonLinGalDyn}), it is also possible
to perform a short integration in physical space, transform the resulting trajectory
into DMAP space, and use the transformed trajectory to {\it estimate} $d \vec L/dt$ ``on demand.''
This is reminiscent of the ``Galerkin-free" computations of \cite{sirisup}, where
this approach was exploited to avoid the construction of the right-hand-side of
a POD-Galerkin dynamical system.
In that case, short bursts of physical simulation observed in POD space were used to
estimate the time-derivatives of the POD components of the solution; these provided
the input to {\it projective} integrators (see, e.g., \cite{man1,gearcoarse}).

Having extrapolated the POD component values to future times, it is easy to construct
physical initial conditions consistent with these values.
In our case, however, having extrapolated
the DMAP solution coordinates to future times, it is more difficult to
find off-sample physical initial conditions consistent with these extrapolated values;
our inverse Nystr\"om map ($\vec \Psi^{-1}$) relied on polynomial interpolation.

Our work started with an available ensemble of points on the manifold, without any
sense of the trajectories from which these points were sampled; if we have the
trajectories themselves, as well as time-derivatives (evaluated or estimated) at
every sample point, then adaptive tabulation tools (see, e.g., \cite{ISAT}) can again help
circumvent the cumbersome computation of the right-hand-side of equation (\ref{nonLinGalDyn}).

We believe that manifold-learning techniques of the type we discussed here, and
the reduced models they lead to, may be particularly useful for tasks for which
small dimensionality is crucial (such as the approximation of stable manifolds or
visualization of the dynamics).

\section{Acknowledgments}

I. G. K. and A. S. are pleased to acknowledge motivating discussions with Dr. R. Erban and Prof. M. S. Jolly.

\bibliographystyle{siam}
\bibliography{NonLinGalManuscript}

\end{document}